# Poking a Dimple in a Black Hole Shows Explicitly that Black Hole Complementarity Violates Causality

Moshe Rozenblit[1]

New York Quantum Theory Group

October 20, 2017

**ABSTRACT**

A massive ball at a fixed distance just outside a black hole (BH) pokes a dimple in the BH by locally depressing the apparent horizon. Analysis of the effect of the dimple on the event horizon shows that if BH Complementarity (BHC) is valid then it is possible to inscribe any message on the surface of the BH; while the message disappears as soon as it is written it can be read by anyone outside the BH during a finite, yet potentially extended period before it is written, thereby violating causality.

# 1 Saving Alice

Alice and Bob exploring a super massive black hole (BH). They are in a space station about 300,000 km outside an eternal Schwarzschild BH with mass M of about a billion solar masses and a Schwarzschild radius R=2GM (in units where the speed of light in vacuum is 1) of about three billion km. G is Newton's gravitational constant. The Schwarzschild shell is defined here as the geometrical locus of points at spatial distance R from the center of the BH. The space station, Alice and Bob and all movements are along a single straight line through the center of the BH. Alice and Bob are each equipped with a standard clock, identical to the stationary clock at infinity that provides the Schwarzschild time coordinate.

At time t=0 on both Alice's and Bob's clocks Alice enters a small spaceship and descends to about 2 mm from the Schwarzschild shell. She gets there at time t=100 seconds in her own proper time. At time t=1,000 seconds on his clock Bob sends a message to Alice to proceed with phase 2 of their planned experiment. As soon as Alice receives the message she turns off her spaceship's engines and descends inertially into the BH. As soon as she crosses the BH apparent horizon she immediately reactivates her spaceship's engines to slow her descent into the BH to the limit allowed by the spacetime curvature.

Some 10 seconds on his clock after Bob sends the message to Alice he lowers a ball of tungsten of mass m of about trillion tons to a few cm from the BH Schwarzschild shell, just above Alice. The ball's descent, assisted with powerful rockets, takes 10 seconds of the Ball's proper time. The BH pull on Alice is counter balanced to a very minute extent by the massive tungsten ball, just enough to allow (a very petite) Alice to escape the BH. The tungsten ball has slightly depressed the apparent horizon, pushing it to a point behind Alice, effectively poking a dimple in the BH apparent horizon, as illustrated schematically in Figure 1 and further examined in Appendix 1.

---

[1] moshe_rozenblit@yahoo.com

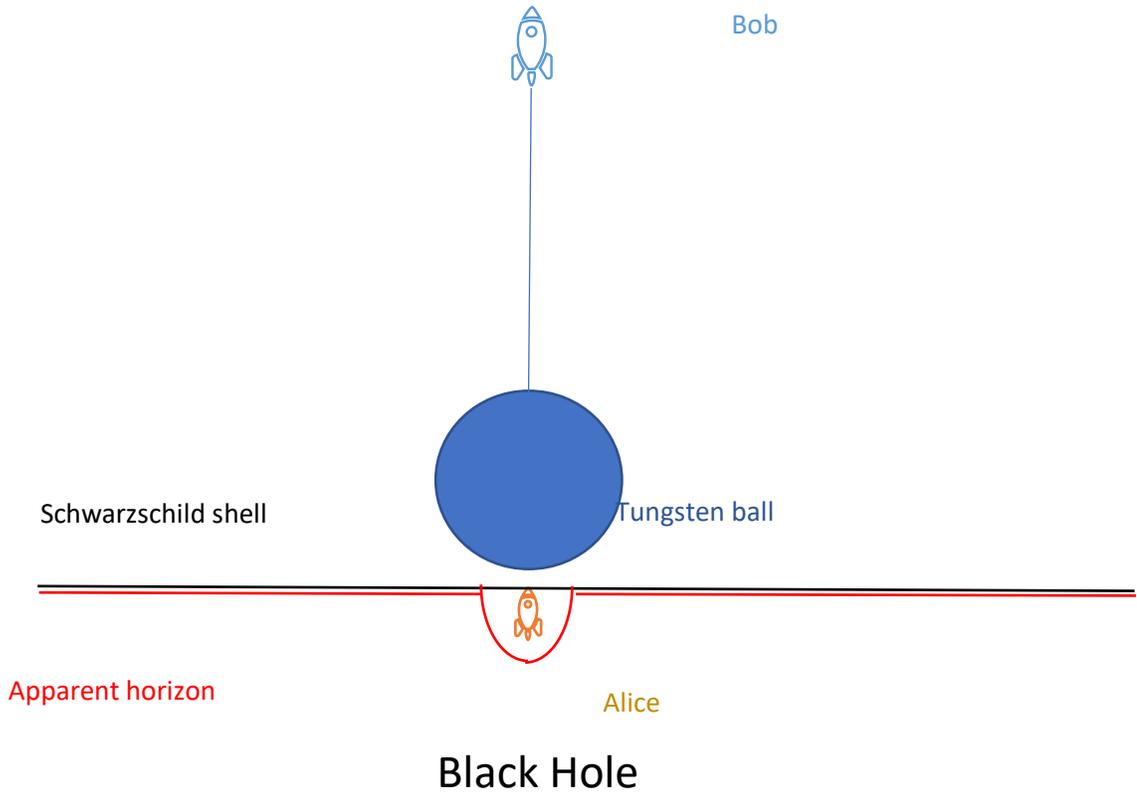

Figure 1: Schematic illustration of a tungsten ball poking a dimple in a BH apparent horizon

It seems that Alice is coming back after having crossed the event horizon, which would be an oxymoron. The fact that Alice escaped the BH means she never crossed the event horizon. The dimple in the apparent horizon also pushed back the event horizon, only more so. Indeed, since Alice did cross the apparent horizon but never crossed the event horizon, the latter must have been depressed before Alice crossed the apparent horizon, let alone before Bob lowered the tungsten ball and poked a dimple in the apparent horizon, as illustrated schematically in Figure 2.

Notice that both figures are very much not drawn to scale; the event horizon, apparent horizon and Schwarzschild shell coincide exactly except where the apparent horizon dips below the Schwarzschild shell and the event horizon dips teleologically below the apparent horizon; furthermore, in Figure 2 the cable connecting the ball to Bob is omitted.

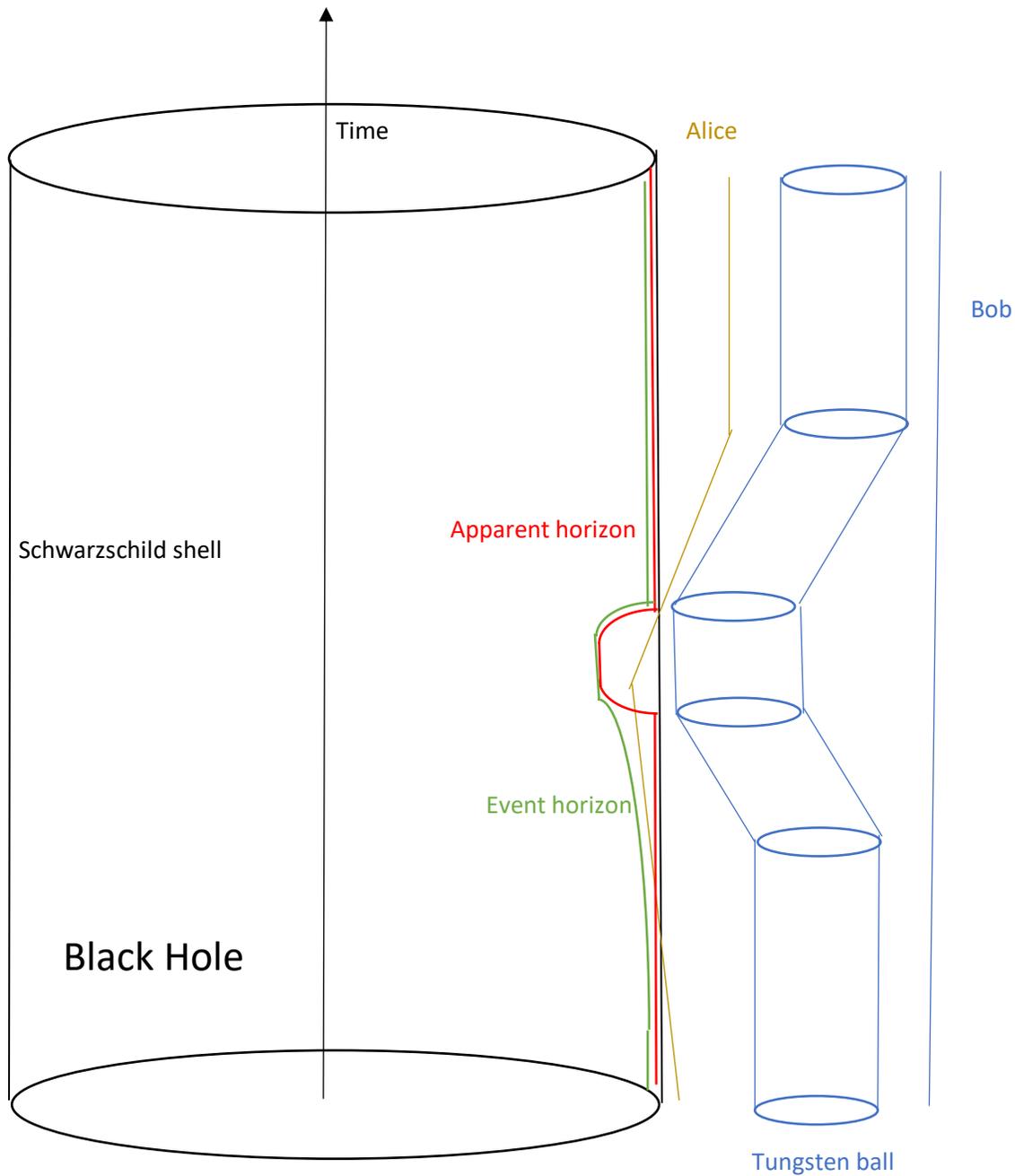

Figure 2: As the tungsten ball pokes a dimple in the BH apparent horizon allowing Alice to escape, the event horizon dips teleologically below the apparent horizon

The fact that the event horizon is depressed teleologically, before the event that caused it (the apparent horizon dimple) happened need not worry us. Indeed, the event horizon is a mathematical construct rather than an observable physical entity or event.

## 2 Writing for the past

BH Complementarity (BHC) [1] has been the most popular vehicle deployed to address the BH information paradox. BHC postulates that the event horizon is enveloped by an extended horizon, about one Planck unit larger than the event horizon itself, and that all the information that would have been inside the event horizon is encoded in this extended horizon. According to BHC the extended horizon is endowed with measurable physical attributes such as entropy, temperature, viscosity, electric resistance and a thermal atmosphere. Since Einstein's Equivalence Principle stipulates that an inertial observer does not detect anything unusual when crossing the event horizon BHC postulates that the stretched horizon cannot be detected by anyone crossing that horizon. This seems to be in contradiction with the measurable physical attributes of the extended horizon listed above, as seen by an observer that remains outside that horizon. BHC reconciles the two divergent claims by noticing they are obtained by two observers who can never communicate and hence compare notes that would show conflicting results.

We are considering two alternatives for what happens to the extended horizon when a dimple is poked in the apparent horizon: (1) the extended horizon remains just outside the Schwarzschild shell or at least outside the apparent horizon, and (2) the extended horizon clings to the event horizon and dips with it below the apparent horizon.

If **the extended horizon remains just outside the Schwarzschild shell or the apparent horizon** then we see from Figure 2 that after Alice falls in through the stretched horizon she manages to emerge from it during her rescue and can then compare notes with Bob. Alice will tell Bob she observed nothing when crossing the Schwarzschild shell while Bob has observed there some significant physical phenomena. Their disagreement would point to irreconcilable internal contradictions within BHC.

If on the other hand **the extended horizon clings to the event horizon** and dips with it below the apparent horizon then this dip occurs well before Bob started lowering the ball to induce a dimple in the apparent horizon. Unlike the teleological dip of the event horizon the dip of the physical extended horizon is clearly detectable and measurable. It is seen by an observer that remains just outside the BH Schwarzschild shell as disappearance of the extended horizon's measurable attributes, such as thermal atmosphere, at the spot where the dimple occurs, but starting well before the dipping of the apparent horizon, when the event horizon starts to dip below the apparent horizon. The blacking out of the stretched horizon's measurable attributes continues until Bob retrieves the ball and the dimple vanishes. Notice, however, that in as far as Hawking radiation originates outside the apparent horizon, independently of the extended horizon, there will be no impact on such perceived Hawking radiation. Nevertheless, any radiation emanating directly from the extended horizon due to its temperature and entropy will be red shifted, dimmed and eventually blacked out as the extended horizon dips below the apparent horizon.

The BH and ball constitute cosmological stationary: Bob can write any message on the surface of the BH by moving the ball over that surface, and Bob, or anyone else who remains just outside the Schwarzschild shell can start reading that message well before it is written, thereby violating causality.

The qualitative arguments above demonstrate that BHC is either internally inconsistent or violates causality, and therefore, in either case, BHC cannot be valid. Appendix 2 provides further support by estimating approximately the time interval before the creation of the dimple that the event horizon starts to dip, as a function of the depth of the dimple.

## Acknowledgements

This paper has been spurred, in part, by valuable critique by Prof Massimo Porrati of an earlier paper. I have also benefited from discussions with members of the NY Quantum Theory Group, in particular with Dr. Andre Mirabelli.

## REFERENCES

[1] L. Susskind, L. Thorlacius, and J. Uglum, Phys. Rev. D **48** 3743 (1993). L. Susskind, arXiv:1208.3445 (16 Aug 2012).

### APPENDIX 1

We do not have an exact solution for Einstein's equation for our specific configuration (BH, ball, Alice, Bob) so by necessity we resort to guesses and approximations.

In Newtonian mechanics gravitational forces are additive – the gravitational force on a test particle due to two massive objects is the sum the two separate forces. The same is true for gravitational potential. In general relativity (GR) gravitational forces are implicit in the metric tensor which determines the geodesics of spacetime. The Schwarzschild metric for the BH is given by

$$ds^2 = \left(1 - \frac{2GM}{r}\right) dt^2 - \left(1 - \frac{2GM}{r}\right)^{-1} dr^2 - r^2 d\theta^2 - r^2 \sin^2\theta d\varphi^2$$

in units where the speed of light in vacuum c=1, *t* is the Schwarzschild time of a standard clock at rest at spatial infinity, and the Schwarzschild radial coordinate *r* is defined so that the area of the 2-sphere at r is $4\pi r^2$. It differs from the flat space Minkowski metric by the term -2GM/r in the coefficients of $dt^2$ and $dr^2$. If the only object around was the ball then a similar expression would apply outside the ball with M and r replaced by the corresponding parameters for the ball – m and $r_b$.

While a ball weighing one trillion tons and 20 times as dense as water may seem imposing by everyday standards, physics in its vicinity is essentially Newtonian. The BH surface gravity is 1/4M which in our case, with M=$10^9$ solar masses is rather mundane. We therefore guess, based on the additive nature of Newtonian forces, that the metric outside the ball, along the line from the center of the BH to the center of the ball, just inside the BH Schwarzschild shell can be reasonably approximated by

$$ds^2 = \left(1 - \frac{2GM}{r} + \frac{2mG}{rb}\right) dt^2 - \left(1 - \frac{2GM}{r} + \frac{2mG}{rb}\right)^{-1} dr^2 - r^2 d\theta^2 - r^2 \sin^2\theta d\varphi^2$$

Thus the apparent horizon, where time becomes space-like, is defined by

$$\left(1 - \frac{2GM}{r} + \frac{2Gm}{rb}\right) = 0$$

2GM/r is thus larger than its Schwarzschild value of 1 by 2Gm/rb which is about 0.7x10$^{-13}$ (after we reinstate c$^2$ in the denominator) in our example; therefore, r is smaller than 2GM (which is about 3 billion km in our scenario) by a similar fraction or about 20 cm. A dimple depth D=20cm does not seem very impressive; however, D can be made much larger by increasing M by a few orders of magnitude and/or replacing the tungsten ball with a solar mass (and earth-size) white dwarf.

# APPENDIX 2

This Appendix estimates the extent of BHC's causality breaking time interval T – how long before the creation of the dimple the event horizon starts to dip as a function of the depth of the dimple. T is the time it takes a photon just one Planck length inside the Schwarzschild shell, (i.e., inside and as close as possible to the Schwarzschild shell) just below the point where the ball comes closest to the BH, and directed radially towards the outside of the BH, to fall a distance D which is the maximum depth of the apparent horizon dimple.

While we do not have an exact solution for Einstein's equation for our specific configuration (BH, ball, Alice, Bob) we use the Schwarzschild metric as a good enough approximation since in the current example the mass M of the BH is larger than the mass m of the ball by about $2 \times 10^{24}$.

We thus start with the Schwarzschild metric

$$ds^2 = \left(1 - \frac{2MG}{r}\right) dt^2 - \left(1 - \frac{2MG}{r}\right)^{-1} dr^2 - r^2 d\theta^2 - r^2 \sin^2\theta d\varphi^2$$

For photons flowing radially from the origin we have:

$$ds = d\theta = d\varphi = 0$$

Thus, the time $T$ to fall a distance D is given by:

$$T = \int_0^T dt = \int_{R1}^{R2} \frac{dr}{\left(1 - \frac{2GM}{r}\right)}$$

Where $R1 = R - l_p$, $R2 = R - D$, $l_p$ is the Planck length and $R = 2GM$.

Thus:

$$T = D - l_p + 2GM \log(D/l_p)$$

In our scenario T~10,000s (after reinstating the factors of c in the denominator); perhaps not very impressive, but undisputable violation of causality nonetheless. Furthermore, in as far as there are no theoretical limits on M and D there is similarly no bound on the extent that BHC can violate causality with a single BH dimple. A determined communicator can further arrange for robotic transcribers at intervals just less than T from any instant in the future to the present.